# STATUS OF THE NSCL CYCLOTRON GAS STOPPER*

N. Joshi[#], G. Bollen, M. Brodeur, D.J. Morrissey, S. Schwarz  
NSCL/MSU, East Lansing, MI 48823, USA

*Abstract*

A gas-filled reverse cyclotron for the thermalisation of energetic beams is under construction at NSCL/MSU. Rare isotopes produced via projectile fragmentation after in-flight separation will be injected into the device and converted into low-energy beams through buffer gas interactions as they spiral towards the centre of the device. The extracted thermal beams will be used for low energy experiments such as precision mass measurements with traps or laser spectroscopy, and further transport for reacceleration. Detailed calculations have been performed to optimize the magnetic field design as well as the transport and stopping of ions inside the gas. An RF-carpet will be used to transport the thermal ions to the axial extraction point. The calculations indicate that the cyclotron gas stopper will be much more efficient for the thermalisation of light and medium mass ions compared to linear gas cells. In this contribution we will discuss simulations of the overall performance and acceptance of machine, the beam matching calculations to the fragment separator emittance, and the construction status.

## INTRODUCTION

The fragmentation of fast heavy-ion projectiles enables fast, chemistry-independent production, separation and delivery of exotic isotopes. The resulting beams of exotic nuclei have high energies (>50 MeV/u) and large emittances due to the production process. The high energy ions are passed through isotope separators followed by a momentum compressor and wedge degrader. These fast ions are slowed down by solid degraders before injection into the gas cell [1].

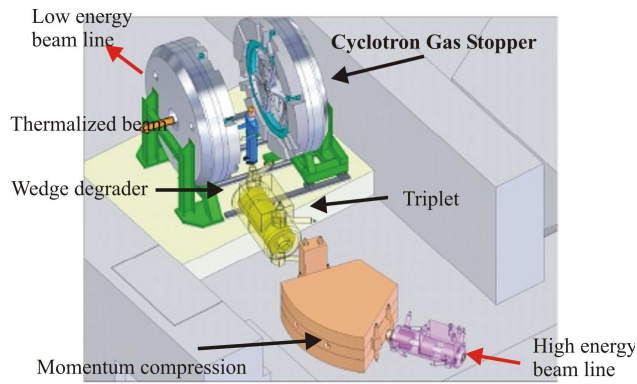

Figure 1: Layout of beam line depicting cyclotron gas stopper in N4 vault of NSCL experimental hall.

The thermalisation and extraction processes in linear gas cell are limited due to large stopping range for light ions and space charge created during the slowing down process. An alternate approach consists of applying a strong axial magnetic field that forces the ions to follow spiral trajectory in the gas. Thermalised ions are then transported using an RF-carpet to the central extraction orifice. Afterwards these ions are transformed into low energy beams using a differentially pumped ion guide and transported for either low energy experiments or for reacceleration. Initial calculations for stopping in such device were presented in [2]. Here, we present simulation results with improved models for stopping, optimisation of the parameters and report the current status of the project.

## BEAM STOPPING AND CYCSTOP CODE

The energetic ions are injected on the outer radius of the cyclotron. Due to the interaction with the buffer gas, the ions lose energy and follow a spiral path inward. The radius ρ of a certain ion is given by expression:

$$\rho = \frac{p}{Bq} = \frac{\sqrt{2mE}}{Bq}, \qquad (1)$$

where p is the momentum, B the magnetic field, q is the charge state, m the mass and E energy of the ion.

The ion dynamics was simulated using a code named CycSTOP. In addition to particle tracking in the magnetic field, the code includes:

- ATIMA module: for the interaction of ion beam with solids.
- SRIM routine: for energy loss of ions in the buffer gas.
- CX package: for the charge exchange process taking place in gasses and solids.
- SAS package: for small angle multiple scattering in gasses.

Fig. 2 shows a model of the cyclotron gas stopper including a magnetic field map and calculated beam envelopes. The energetic beam (magnetic rigidity Bρ=2.6 Tm) after injection passes through a vacuum window, made of a metal foil e.g. Al or Be; before it hits the glass degrader which brings the energy of ion down to match a magnetic rigidity of 1.6 Tm. The ions spiral inward towards the centre and are considered as stopped once the energy falls below the cut off value of 1 keV. The ions are regarded as lost if they hit the injection channel, the axial or radial wall, or stay in the degrader.

The CycSTOP code has been upgraded to describe the slowing down process more realistically since the original

___________________________________________  
*Work supported by NSF  
#joshi@nscl.msu.edu

work described in Ref. [2] with the following improvements:

- The tables of charge exchange cross sections have been rebuilt. At high energies the ETACHA code [3] was used to calculate the single electron capture and loss cross sections. At low energies Schlachter's empirical scaling rule [4] was used to calculate the single electron capture and equilibrium charge state was calculated using Schiwietz's formula [5]. The electron loss cross sections are based on Franzke's method [6]. The high and low energy results were joined by interpolation.
- Small angle multiple scattering [7] has been added and tested against data available in the literature. The CycSTOP code was modified to allow gas targets with arbitrary thickness and atomic number.
- The particle mover has been changed to account for relativistic motion of ions with the implementation of the Boris method [8]. The relativistic γ factor typically lies in the range of 1.05-1.07 for the injection energies. The relativistic and non-relativistic calculations were found to differ in stopping efficiency in range of 3-5%.

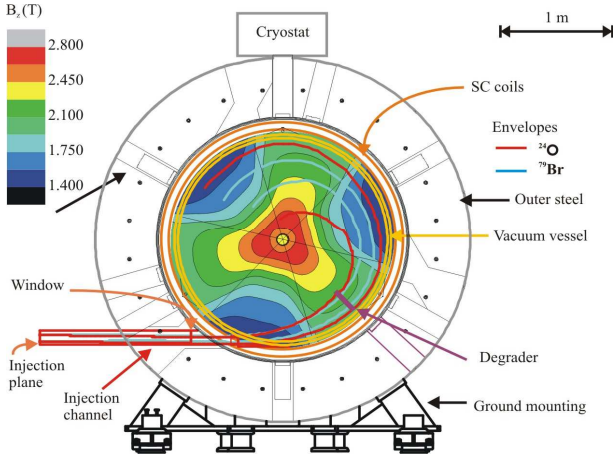

Figure 2: The cut-view model of the cyclotron gas stopper. The magnetic field map is overlaid along with beam envelope for to ion species $^{24}$O, $^{79}$Br.

The sectored magnet consists of hill and valley profile with 3 fold symmetry to provide axial focussing. The magnetic field with a maximum strength of about 2.7 T will be generated by a superconducting coil excited by currents up to 320 kA. The magnetic field configuration was optimized in order to achieve maximum stopping efficiency. The magnet discussed in [9] (version S13) required further modification to gain a higher acceptance for lighter ions.

To calculate the acceptance, the ions were distributed at the injection plane (about 2 m outside) with all possible positions and angles that correspond to a homogenous distribution in the phase-space. The ions which successfully stopped were sampled to get the acceptance for a particular configuration. Fig. 3 shows an example of such acceptance plot for $^{79}$Br ions calculated at a gas pressure of 100 mbar.

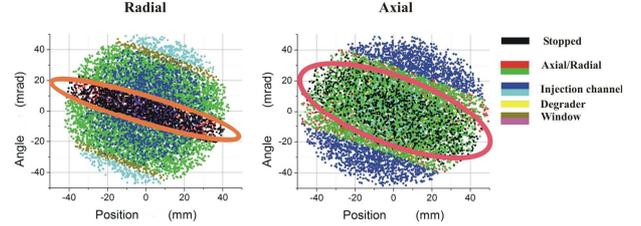

Figure 3: Example of an acceptance plot for $^{79}$Br ions at a gas pressure of 100 mbar. The phase-space is colour coded according to the fate of ions.

Table 1 compares the results for the two magnet designs; the S13 with a pole gap of 75 mm and the S17 with a gap of 90 mm, in terms of acceptances at a gas pressure of 100 mbar. It can be seen that the axial acceptance ($\varepsilon_z$) is considerably increased for $^{24}$O, by about 23%. No significant change was found in radial phase-space ($\varepsilon_r$).

Table 1: Acceptance comparison for two versions of magnet for different ion species. All values have units of π-mm-mrad

| Ion | $\varepsilon_r$ (S13) | $\varepsilon_z$ (S13) | $\varepsilon_r$ (S17) | $\varepsilon_z$ (S17) |
|---|---|---|---|---|
| $^{79}$Br | 459 | 909 | 510 | 982 |
| $^{56}$Fe | 466 | 889 | 489 | 969 |
| $^{40}$Si | 495 | 907 | 503 | 1044 |
| $^{24}$O | 506 | 797 | 501 | 1038 |

A few more iterations of the magnet design were performed for pole profile to reduce axial losses contributed by Walkinshaw resonance. The magnet field was finalised with a version called S20.

The beam optics calculation for beam passing through the A1900 separator was done using combination of two well known codes TRANSPORT and LISE++ [10,11]. The beam emittances from LISE++ were matched to the calculated acceptances. Table 2 gives the axial and radial acceptances for the S20 magnet and emittances calculated by LISE++ for selected ion species. The calculations show promising efficiencies even for lighter ions like $^{24}$O at a low gas pressure of 100 mbar.

Table 2: Acceptance for S20 magnet, input emittances for different ions calculated from LISE++ and corresponding efficiency. All values have units of π-mm-mrad

| Ion | $\varepsilon_r$ (S20) | $\varepsilon_z$ (S20) | $\varepsilon_r$ (LISE) | $\varepsilon_z$ (LISE) | Efficiency (%) |
|---|---|---|---|---|---|
| $^{79}$Br | 897 | 1190 | 227 | 424 | 98.10 |
| $^{56}$Fe | 740 | 1165 | 153 | 419 | 96.85 |
| $^{40}$Si | 853 | 1187 | 336 | 1098 | 86.50 |
| $^{24}$O | 707 | 1179 | 1550 | 1038 | 64.90 |

The energy spread acceptance for the cyclotron gas stopper is dominated by the interaction with degrader. Ions moving too slowly are lost inside the degrader and ions with too high energy are lost radially due to a rigidity mismatch. The energy spread acceptance is asymmetric ranging typically from -10% to +5%. The expected energy spread calculated with LISE++ is symmetric with a Gaussian distribution. Table 3 compares the typical values for energy acceptance and values obtained with LISE++ after momentum compressor.

Table 3: Longitudinal energy acceptance for S20 magnet and energy spread calculated from LISE++.

| Ion | $^{79}$Br | $^{56}$Fe | $^{40}$Si | $^{24}$O |
|---|---|---|---|---|
| ΔE/E (S20) | -10.6% ~ +5 % | -10.1% ~ +2.1 % | -10.1% ~ +2.8 % | -9.5% ~ +0.9% |
| ΔE/E (LISE) | ± 1.3 % | ± 1.3 % | ± 1.4 % | ± 4.6 % |

Fig. 4 shows variation in the stopping efficiency as a function of gas pressure. It can be seen that the cyclotron gas stopper can even be operated at low gas pressure in case of heavier ions, whereas in case of lighter ions at a nominal pressure of 100 mbar it exhibits as efficiency of more than 65%, which can be further increased by increasing in the gas pressure to 150 mbar.

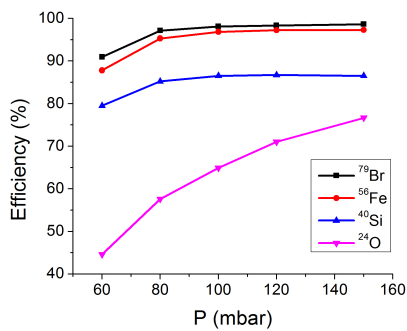

Figure 4: Stopping efficiency as a function of buffer gas pressure for different ions.

## RF-CARPET

The ions, once thermalised, need to be transported radially from the outer peripheral region towards a central aperture for extraction. An RF-carpet will be employed for this task. The carpet will be driven in so-called ion-surfing mode [12], which replaces the drag field with a travelling wave. Experiments were performed using two different isotopes. $^{85}$Rb$^+$ ions were successfully transported over a distances up to 40 cm with gas pressures reaching 240 mbar. For a pressure of 120 mbar, $^{85}$Rb$^+$ ions were shown to sustain axial push fields of over 45 V/cm, and reached transport velocities of 75 m/s. Efficient transport of $^{39}$K$^+$ ions was also achieved at a pressure of 80 mbar with an axial push field of 20 V/cm and velocities of 50 m/s.

## CURRENT STATUS

Detailed ion stopping simulations have been performed to optimise acceptance of the cyclotron gas stopper. The magnet design is finalized. The project has entered into the design and manufacturing phase. The pole pieces were ordered are being machined and the yoke steel is being manufactured. Considerable amount of effort are being devoted to designing vacuum vessel, cryostat and mountings. The whole assembly is scheduled to be tested around middle of 2013. The scaled up version of the RF-carpet and the extraction system using ion guide are being designed. Both of these will be tested separately before final installation into the gas stopper. The final commissioning of the cyclotron gas stopper is expected late in 2013.

## ACKNOWLEDGMENT

The project is supported by the National Science Foundation under grant PHY-09-58726.